\documentclass[12pt]{iopart}
\usepackage{graphicx}
\begin{document}
\title{Following microscopic motion in a two dimensional glass-forming binary fluid}
\author{Matthew T. Downton$^{1,2}$ and Malcolm P. Kennett$^{1}$}
\address{1 Physics Department, Simon Fraser University, 8888 University
  Drive, Burnaby, British Columbia, V5A 1S6, Canada}
\address{2 Institut f\"{u}r Theoretische Physik, Technicshe Universit\"{a}t Berlin, 10623 Berlin, Germany}

\begin{abstract}
The dynamics of a binary mixture of large and small discs are studied at temperatures
approaching the glass transition using an analysis based on the topology of the
Voronoi polygon surrounding each atom. At higher temperatures we find that dynamics
is dominated by fluid-like motion that involves particles entering and exiting the
nearest-neighbour shells of nearby particles.  As the temperature is lowered, the
rate of topological moves decreases and motion becomes localised to regions of mixed pentagons
and heptagons. In addition we find that in the low temperature state particles may 
translate significant distances without undergoing changes in their nearest neighbour shell.
These results have implications for dynamical heterogeneities in glass forming liquids.
\end{abstract}

\maketitle

\section{Introduction}
\label{sec:introduction}

The relationship between the dynamics and structure of a 
glass-forming liquid is a non-trivial problem.
Structurally, a simple liquid is isotropic
and has smooth pair distribution functions that decay to the 
background value within a few nearest neighbour spacings. The positions of the maxima and 
minima of these functions can be explained in terms of coordination shells of particles 
packed around a central reference particle.  This gives an intuitive understanding of the 
microscopic structure of a liquid. The corresponding picture of dynamics is based
around atoms that are trapped by their surrounding neighbours, and make rare ``cage-breaking''
jumps to nearby positions. At high temperatures, the separation in time between collisions
with neighbours and jumps is small, but at lower temperatures this time starts to diverge.
How different arrangements of atoms promote or inhibit this motion is an open question.

Molecular dynamics simulations provide a powerful tool to address this problem by
allowing one to follow structure and dynamics simultaneously.  It has been observed
in simulations that systems that exhibit glass-like behavior at lower temperatures 
can have characteristics of an interplay between crystal and liquid structure 
\cite{shintani06}. Structural changes have also been observed in fluids 
approaching the fluid-solid transition \cite{truskett98}.  Additionally, it
has been observed that as the glass transition is approached, dynamics become
increasingly spatially heterogeneous 
\cite{Sillescu,Ediger,Glotzer,Richert,Castillo,Israeloff,VandenBout,Reinsberg,Weeks,Kob,Donati1,Donati2}.  
The character of these dynamical heterogeneities has been well studied,
both computationally \cite{Donati1,HarrowellDH,Yamamoto,Andersen,Perera2,Gebremichael1,Stevenson} 
and in experiments on colloidal glasses  \cite{Kegel,Weeks2}, in which ``caging'' of particles,
and the escape from cages via string-like motion has
been seen at temperatures slightly above the glass transition temperature.  
Despite all of this effort, the origin of dynamical heterogeneities is unclear, as
is their connection to the dramatic slowing down of dynamics seen as a liquid becomes glassy.

A simulation approach that has been recently introduced, the iso-configurational ensemble, 
sheds light on the relationship between structure and dynamics \cite{widmer-cooper04}. 
In this technique, repeated short simulations are performed using the same starting structure, but 
randomized velocities.  Averages of single particle displacements over the 
ensemble of simulations give a map of the propensity to move.  It has been observed that
 the local Debye-Waller factor appears
to correlate very closely with the dynamic propensity at long times \cite{widmer-cooper06},
however it is harder to
answer the more subtle question of how the propensity correlates with the dynamics
of a single trajectory.  It appears that on the single-particle level, such a connection
cannot be made \cite{berthier07}, but that the connection between structure and dynamics
will exist at longer lengthscales.

An alternative theoretical approach to understanding the behaviour of glass-forming liquids, 
and in particular their kinetic arrest in the vicinity of the glass transition 
temperature, has been through studying phenomenological models.  Such models, 
in particular, kinetically constrained models (KCMs) have received considerable
attention in recent years 
\cite{ritort03,FA84,FA85,Graham97,Garrahan02,Berthier03,PNAS03,Leonard07}
as they have had success in reproducing diverging timescales observed in glass-formers.
These models are generally understood as an effective model for some coarse-grained
degree of freedom (often referred to as spins), and at low temperatures, the dynamics can 
be understood as a few up-spins or ``defects'' in a background sea of
down spins \cite{BerthierPRE03}.  We have recently attempted to provide a
coarse-graining procedure to map the dynamics of a glass-former onto a KCM, using
local mean squared displacement as a means to define spins \cite{Downton07}.
This approach found kinetically constrained behaviour became increasingly important
at low temperatures, but only on timescales longer than the alpha relaxation time,
whereas one would hope that the alpha relaxation time was in fact a consequence of the
KCM, rather than an input.
For shorter coarse-graining times, there was no evidence of kinetically constrained
dynamics.  This suggests that alternative defect variables should be identified if
such a coarse-graining procedure is to be successful.  

In this paper we examine the interplay between structure and dynamics 
by investigating the dynamics of  a two-dimensional liquid in terms of the 
coordination number given by a Voronoi polygonisation. Previously, 
the population densities of different coordinations
have been used to make predictions regarding the glass transition temperature \cite{aharonov07}.
Here we use this choice to give an unambiguous definition of `fluid-like'
motion within a dense particle system in terms of a simple topological process.
The approach we take can also be related to previous work on lattice based models
that obey a similar dynamical rule \cite{davison00,davison01,sherrington02}.
We find that there are a few dominant particle moves, of which, only one
has significant temperature dependence, becoming relatively more 
important with decreasing temperature.

The structure of the paper is as follows:  in Sec.~\ref{sec:model} we give
details of the model and simulation procedures, 
along with some results that show the development of glassy phenomenology 
at lower temperatures.  
In Sec.~\ref{sec:results-simulations}
we define the ``defects'' we consider and discuss their spatial distribution,
and in Sec.~\ref{sec:topol-based-dynam} we present the microscopic dynamics of the fluid
in terms of local topology changing transitions.  
 These results are followed by concluding observations and discussion in 
Sec.~\ref{sec:discussion}.

\section{Model and simulation details}
\label{sec:model}

We used a two-dimensional model that exhibits all of the key ingredients of
a true glass-former. The model consists of a binary mixture of large and small soft 
discs with size ratio 1:1.4 interacting with a potential of the form 
$V_{\alpha\beta}=\epsilon \left(\frac{\sigma_{\alpha\beta}}{r}\right)^{12}$. 
The mass, $m$, of all particles was set to unity; time was measured in units 
$\tau = \sqrt{m\sigma_{11}^{2}/\epsilon}$; distance was measured in terms of the
small particle diameter, $\sigma_{11}$; temperature was measured in units of $\epsilon$.

Molecular dynamics simulations of 1600 particles in the ratio 75:25 small to large were performed
at constant volume and temperature with the number density fixed to 
$0.85\sigma_{11}^{-2}$.  Configurations were equilibrated at 
high temperatures then slowly cooled to create a series of configurations at 
different temperatures. These were then extensively equilibrated at fixed 
temperature using a Gaussian isokinetic thermostat. 
At temperatures below $T=0.330$ the system was found to age as indicated by a continuous
decrease in the pressure and total potential energy.
In a separate series of calculations it was found that when the initial 
configuration consisted of segregated small and large particles, there
was complete mixing  for temperatures $T\ge 0.450$; below this 
temperature there was no mixing for initially segregated configurations 
other than relaxation at the phase boundary.

\begin{figure}
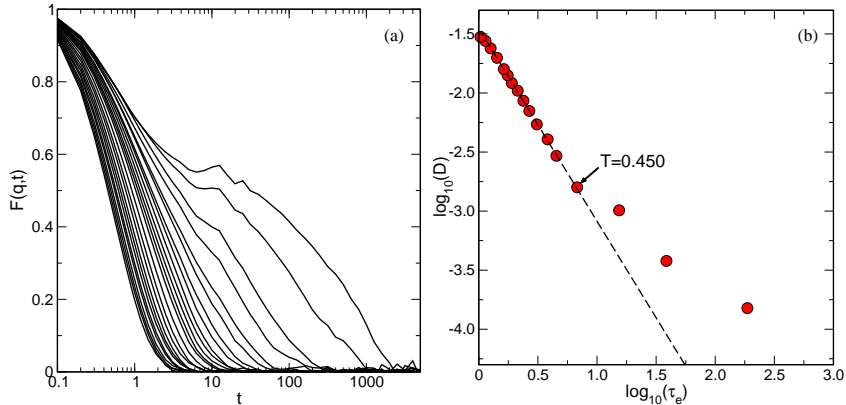

  \centering
  \includegraphics[width=0.35\columnwidth]{figure1a.eps}
  \includegraphics[width=0.35\columnwidth]{figure1b.eps}
  \caption{Development of glassy behavior at lower temperatures. 
(a) The dynamic structure factor, $F(q,t)$, calculated at $q=6.23\sigma^{-1}$ ($T=0.360 - 1.440$) 
(b) Change in scaling of the structural relaxation time, $\tau_{\alpha}$, and the diffusion 
coefficient for small particles.}
  \label{fig:background}
\end{figure}

The phenomenology of this mixture is almost identical to the mixture studied by
Perera and Harrowell: multiple relaxational timescales develop at lower 
temperatures; a change in the scaling between structural relaxation and 
diffusion occurs at a temperature $T^{*}$ \cite{perera98,perera99,perera99a}. 
These results are summarised in 
figure~\ref{fig:background} where we take the structural 
relaxation time, $\tau_{\alpha}$, to be the time it takes the function
\begin{displaymath}
  F(q,t) = \frac{\left<\rho(\mathbf{q},t)\rho(\mathbf{-q},0)\right>}{\left<\rho(\mathbf{q},0)\rho(\mathbf{-q},0)\right>},
\end{displaymath}
 to decay to the value $1/e$, 
where $\rho(\mathbf{q},t) = \sum_{j}\exp\left( i\mathbf{q}\cdot\mathbf{r}_{j}\right)$ 
and $q=|\mathbf{q}|$ is the position of the first peak in the static structure 
factor. Similar results were found if the 
self-intermediate scattering function for a single particle species was used
to measure structural relaxation.  It can be seen that $T^{*}\approx 0.450$ 
for this particular mixture: at temperatures below $T^*$, 
$F(q,t)$ begins to exhibit two stage decay.

\section{Static results}
\label{sec:results-simulations}

\subsection{Distribution and concentration of defects}
\label{sec:distribution-defects}

The Voronoi procedure partitions the system into polygonal cells surrounding each atom. 
Nearest neighbours are identified as atoms which share an edge, whereas next-nearest
neighbours have common nearest neighbours but are not themselves neighbours. 
For convenience we refer to atoms with coordination number, $c$, differing from six
as defects and use the ``topological charge'', $q=c-6$, to characterise the 
coordination of individual atoms. Within simulations such as this where periodic boundary conditions
are used, the total coordination is conserved with $\sum_{i}q_{i}=0$.

\begin{figure}
  \centering
  \includegraphics[width=0.5\columnwidth]{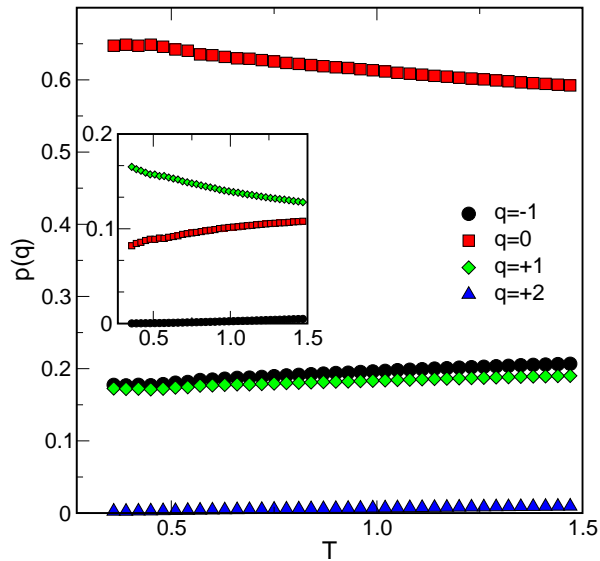}
  \caption{Population of defect species as a function of temperature in the non-aging temperature range. Inset: Topological charge of large atoms as a proportion of the total number of atoms.}
  \label{fig:defect}
\end{figure}

The variation with temperature of the concentration of atoms with different $q$ is
shown in figure~\ref{fig:defect}.  
Although occasional configurations contained four-fold or nine-fold coordinated atoms
these were found to be relatively rare and the majority of atoms have $q$ in the range
-1 to +2.
At higher temperatures there is an excess of $-1$ atoms to $+1$ atoms that is 
compensated for by a small number of $+2$ atoms. At all temperatures within the range of
study, the majority of large atoms have charge $+1$ and these constitute almost all of the +1 atoms
at the lower end of this range (inset to figure~\ref{fig:defect}).  The number of $q=0$ atoms
also increases as a function of $T$, before levelling out at $T^{*}$.

\begin{figure}
  \centering
  \includegraphics[width=0.5\columnwidth]{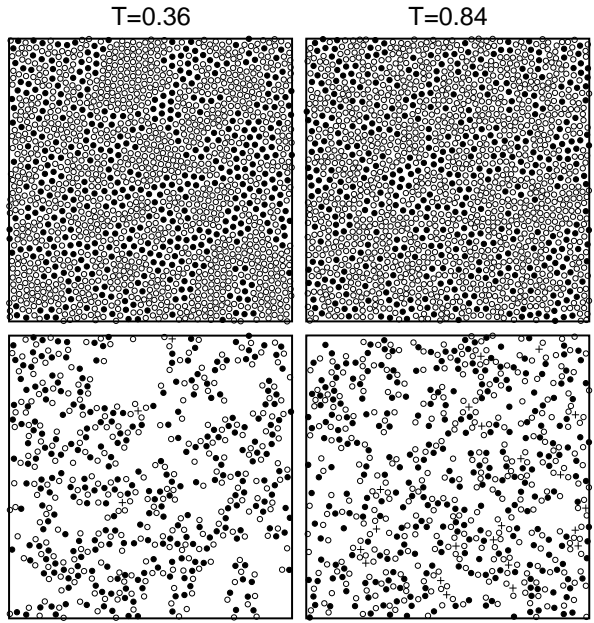}
  \caption{Particle positions and coordination at high and low temperatures. From left to right: $T=0.36$ and $T=0.84$. The top configuration shows the spatial distribution of large and small particles as filled and open circles respectively.  
In the lower configuration, atoms with $q=+1$ are filled circles, atoms with $q=-1$ are open circles, atoms with $q=+2$ are marked with a (+).}
  \label{fig:configurations}
\end{figure}

Two configurations representative of high and low temperatures are shown 
in figure~\ref{fig:configurations}. In both configurations, the distribution of large and small particles
is similar and there is no apparent demixing of the system at lower temperatures. However,
a greater degree of hexagonal coordination is seen within the lower temperature 
configuration; this is especially noticeable in larger domains of small particles.  
Accompanying these snapshots
 are pictures of the system showing the positions of atoms with non-zero $q$. 
For $T=0.870$ there is a slightly higher concentration of lone defects. In both systems
there is a network of defects which almost spans the system.  Inspection of figure~\ref{fig:defect}
shows that in fact at the temperature at which the Stokes-Einstein relation breaks down, 
$T^{*}$, the population of $q=0$ sites reaches $\sim 0.65$, and the threshold for 
bond percolation on a triangular lattice is 0.347 \cite{Sykes}.  Relationships between 
percolation and breakdown of the Stokes-Einstein relation have been noticed by
earlier authors in the context of a lattice gas \cite{Nicodemi} and a three dimensional 
glass former \cite{Dzugutov}.  The defect
network is more tightly connected at lower temperatures, and leaves larger regions that are
defect free and consist of small particles.  Defects are generally arranged in pairs, so that
a +1 defect is the neighbor to a -1 defect; no net separation of charge can be observed at either
high or low temperatures.

\begin{figure}
  \centering
  \includegraphics[width=0.5\columnwidth]{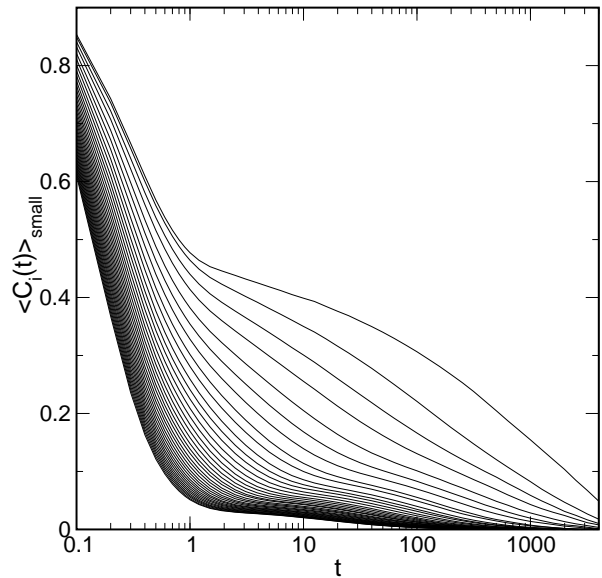}
  \caption{Autocorrelation function of topological charge averaged over small
    atoms. From top: temperatures in the range $T=0.360-1.470$ in
    steps of $0.03$.}
  \label{fig:spin}
\end{figure}

In figure~\ref{fig:spin} we plot the temperature dependence of the autocorrelation function of 
$q$.  This is defined as
\begin{displaymath}
  \label{eq:autocorrelation}
  C_{i}(t) = \frac{\left<q_{i}(t)q_{i}(0)\right>-\left<q_{i}\right>^{2}}{\left<q_{i}^2\right>-\left<q_{i}\right>^{2}}.
\end{displaymath}
At high temperatures $C(t)$ decays rapidly before plateauing and then decaying more slowly.
Reducing the temperature gives rise to a more complicated decay pattern and a second
long-time decay rate appears that grows at a similar rate as the dynamic structure factor, $F(q,t)$.
Hence the dynamics of structural ``defects'' clearly mirrors that of density correlations, 
suggesting that there may be some connection between the two.  

\section{Topology based dynamics}
\label{sec:topol-based-dynam}

\begin{figure}
  \centering
  \includegraphics[width=0.5\columnwidth]{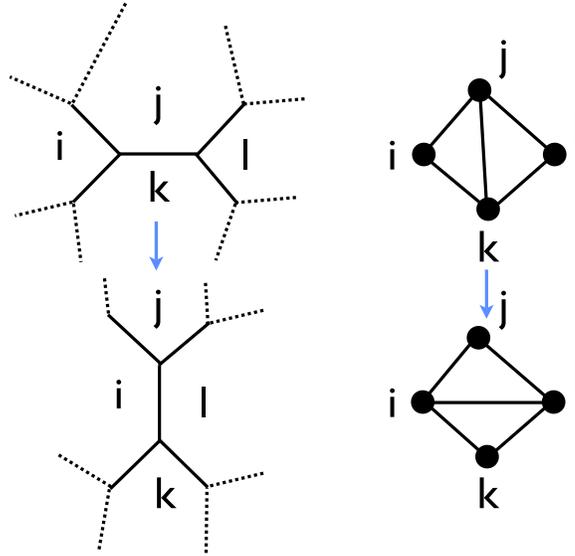}
  \caption{The T1 move in both representations of the particle positions (a) from the Voronoi cells (b) from the Delaunay triangulation of nearest neighbours. In these dual views, $i$ and $l$ move together to become nearest neighbours, $j$ and $k$ are pushed apart, becoming next-nearest neighbours.  All other neighbours of the four atoms remain the same.}
  \label{fig:T1}
\end{figure}

One output of a Voronoi tesselation is a graph of nearest neighbours within the system.
If fluid-like motion in real space can be thought of 
as the random motion of particles between the nearest and 
next-nearest neighbour shells then this motion has a simple topological rule that governs the creation
and destruction of edges 
on the graph or equivalently the changes in the number of edges on each Voronoi
cell.  These dual views of the dynamics are shown in figure~\ref{fig:T1}. 
It is clear from the figure that
pairs of particles that become nearest neighbours each gain an edge from
 the displaced particles. Provided
that the number of particles is conserved, this single rule can be used to describe all of the changes
in the coordination that occur during a simulation or experiment \cite{larson}.  
This gives a suitable coarse-graining of configurational changes and allows us to 
examine the microscopic origin of the decay of the topological charge found in the previous section.

In the following section we proceed
as follows:  A standard molecular dynamics simulation is performed and the Voronoi 
procedure is performed
frequently enough that the majority of nearest neighbour changes can be directly 
identified\footnote{In practice, with a timestep of $\Delta t = 0.0005\tau$, 94\% of T1 moves 
can be identified at $T=0.840$ and 98\% at $T=0.360$. The remaining moves are 
groups of 5-8 atoms that change coordination simultaneously.}.
We then examine the frequency with which moves occur, the type and spatial distribution of 
nearest neighbour changes and finally examine the relationship with particle displacements at 
low temperatures. In the present study we ignore the identity of the atoms.

\subsection{Identity and spatial distribution of nearest neighbour exchanges}
\label{sec:rate-near-neighb}

The majority of nearest neighbour exchanges consist of the six basic moves listed in 
table~\ref{tab:processes}. In this table, we have listed the topological charge of each atom
\textit{before} the nearest neighbour exchange occurs. The atoms in each case are labelled 
according to figure~\ref{fig:T1}: $i$ and $l$ 
are acceptor atoms that each gain an extra nearest neighbour after the T1 move; $j$ and $k$
both lose nearest neighbours after the exchange. Many other exchange types are observed
in the simulations, but are relatively infrequent in comparison to those listed.

\begin{figure}
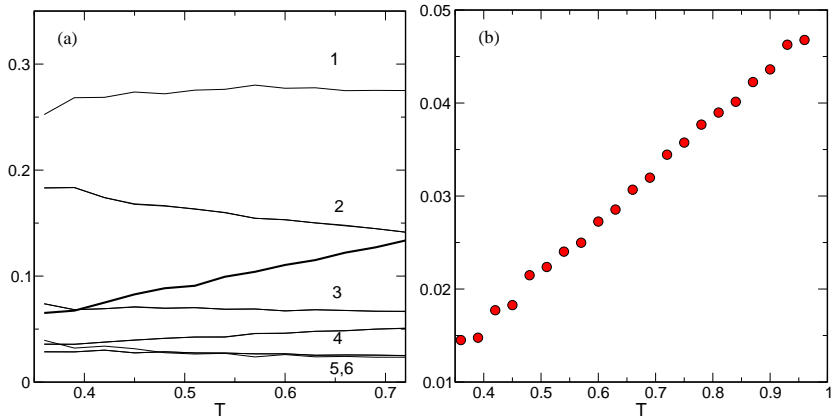

  \centering
  \includegraphics[width=0.34\columnwidth]{figure6a.eps}
  \includegraphics[width=0.35\columnwidth]{figure6b.eps}
  \caption{(a) Temperature dependence of the probability of the six most frequent T1 moves. The labelled curves refer to the moves in listed table~\ref{tab:processes}; the thick solid line indicates the probability of all other moves.  For moves $\mathbf{2}, \mathbf{3}, \mathbf{4}$ and $\mathbf{6}$ both sets of curves are plotted, but are almost indistinguishable from one another. (b) Number of T1 moves per particle per unit time.}
  \label{fig:temp-dependence}
\end{figure}

The temperature dependence of the relative frequency of 
these move types is plotted in figure~\ref{fig:temp-dependence}(a).
Two aspects of the dynamics are immediately apparent.  First, there is no sudden change in the 
relative frequency of any particular T1 move at the temperature where the change in scaling between 
diffusion and structural relaxation occurs.  Second, moves which are the inverse of one another
are found to occur with similar frequency.  This can be seen as an indication that detailed
balance holds for the liquid and that the simulations are in equilibrium on the time scales 
studied. In figure~\ref{fig:temp-dependence}(b), the rate at which T1 moves are observed is plotted 
as a function of temperature and can be seen to be surprisingly linear.

At all temperatures, the most significant move is the dislocation glide (move $\mathbf{1}$) 
where a pair of
$+1$ and $-1$ defects translate by a single particle spacing in a direction that is roughly
perpendicular to the line segment that connects the two atoms.  The frequency of this move
has been noted in previous simulation studies of liquid dynamics \cite{deng89b}.
The move $\mathbf{2}_\mathrm{a}$ involves the combination of a $-1$ defect with three atoms that
have coordination of six to produce two $-1$ defects, a single $+1$ defect and a six-fold coordinated
atom. $\mathbf{2}_\mathrm{b}$ is the reverse of this move.  This pair of moves is therefore one route
by which a lone $\pm 1$ defect pair may be created or annihilated and is the only move of the most frequent
that shows any strong temperature dependence.
The final move that we  consider in detail is the pair $\mathbf{3}_\mathrm{a}$ and 
$\mathbf{3}_\mathrm{b}$.  This involves the destruction of two $+1$ and two $-1$ 
defects in one direction or the creation of these defects in the other direction and will play an 
important role in the melting and annealing of crystal lattices.

\begin{table}
  \centering
  \begin{tabular}{lrrrrrlrrrrr}
    \hline
 & $i$ & $j$ & $k$ & $l$ & $c$ & & $i$ & $j$ & $k$ & $l$ & $c$\\
    \hline
    $\mathbf{1}$ & -1 & 1 & 0 & 0 & 0 & $\mathbf{4}_\mathrm{a}$ & -1 & 1 & 1 & 0 & 1\\
    $\mathbf{2}_\mathrm{a}$ & -1 & 0 & 0 & 0 & -1 & $\mathbf{4}_\mathrm{b}$ & 0 & 0 & 1 & 0 & 1\\
    $\mathbf{2}_\mathrm{b}$ & -1 & 0 & 1 & -1 & -1 & $\mathbf{5}$ & -1 & 0 & 0 & -1 & -2\\
    $\mathbf{3}_\mathrm{a}$ & -1 & 1 & 1 & -1 & 0 & $\mathbf{6}_\mathrm{a}$ & -1 & 0 & 0 & 1 & 0 \\
    $\mathbf{3}_\mathrm{b}$ & 0  & 0 & 0 &  0 & 0 & $\mathbf{6}_\mathrm{b}$ & -1 & 0 & 2 & -1 & 0\\
    \hline
  \end{tabular}
  \caption{Processes that comprise most of the nearest neighbour exchanges observed in the simulations.  The columns $i, j, k$ and $l$ use the same atom labels as figure~\ref{fig:T1} and give the topological charge on each atom \textit{before} the exchange occurs.  The number in bold refers to the label used in figure~\ref{fig:temp-dependence}.  Exchanges which are the mutual reverse of one other are labelled with the subscripts $a$ and $b$.}
  \label{tab:processes}
\end{table}

\begin{figure}
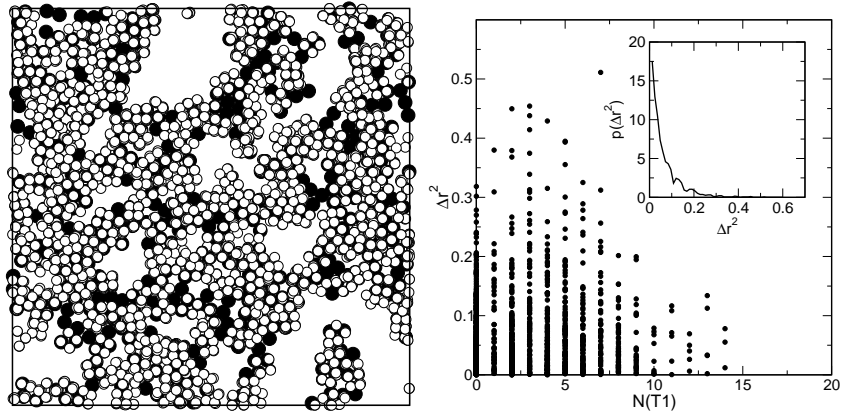

  \centering
  \includegraphics[width=0.35\columnwidth]{figure7a.eps}
  \includegraphics[width=0.35\columnwidth]{figure7b.eps}
  \caption{(a) Distribution of 1000 T1 moves (white circles) at $T=0.360$ superimposed on the positions of the large atoms in the starting configuration (black circles) (b) Correlation between the number of T1 moves that an atom undertakes and the mean squared displacement, $\Delta r^{2}$ (inset) Probability distribution function of $\Delta r^{2}$.}
  \label{fig:moves}
\end{figure}

Having identified moves as they occur, the next stage in understanding the dynamics of the 
liquid is to examine the spatial distribution of nearest neighbour exchanges.  
In figure~\ref{fig:moves}(a),
we show the distribution of the first 1000 T1 moves observed after the $T=0.360$ snapshot shown in 
figure~\ref{fig:configurations} was taken. These are strongly localised around the mixed regions of +1
and -1 defects  that were previously observed in figure~\ref{fig:configurations}. 
Figure~\ref{fig:moves}(b) shows that there is not a strong correlation between the number of T1 
moves observed for a given particle and the distance that it translates.  
In fact there seem to be a number of particles that displace relatively
large distances during the short simulations without any T1 moves.  However, the maximum displacement
for particles that experience large numbers of T1 moves does tend to be smaller than those with 
fewer T1 moves.

One problem with the method that
we present here is that we have no simple way of accounting for quartets of atoms are in almost square 
conformations.  Small fluctuations around particle positions could then lead to the mis-counting of a large
number of T1 moves.  To address this we have repeated the simulations and changed the frequency with which
the Voronoi procedure is performed and do not find any changes in the overall rate of T1 moves. Such counting
problems are therefore unlikely to be significant.

\section{Conclusions}
\label{sec:discussion}

We have presented results of simulations of a two dimensional glass-forming binary fluid that shows
identifiable changes in its structure when the control temperature is lowered. These changes are
related to the spatial distribution of atoms with coordination that differs from the mean value.
Using the topological properties of the two-dimensional nearest neighbour network we have shown that
it is possible to identify a strong relationship between one aspect of the spatial 
distribution of dynamics and the local level of disorder in structure.  There is apparently still 
motion within the simulation cell that is not accounted for by the topological process that 
we study, but our work may contribute to understanding of the puzzle presented by 
heterogeneous dynamics in slow and glassy systems.  The observation that the rate of T1 moves
decreases linearly with temperature, with the relative proportions of different types of T1 moves
fairly insensitive to temperature, suggests that the slowing down associated with the 
glass transition is likely not to be able to be ascribed to any one set of T1 processes.  
Recent work has suggested that a particular class of ``liquid-like'' defects (which presumably
have some set of associated topological moves) may play a role in the glass transition 
\cite{aharonov07}, but we have not explored this possibility here. However,
the temperature dependence of the time decay of the correlations in topological charge, 
and its similarity to the decay 
of the dynamic structure factor suggest that the dynamics of T1 moves may be relevant to
dynamic facilitation as is required for kinetically constrained models of glasses.

\ack

Michael Plischke is thanked for encouragement and helpful discussions. NSERC 
and Westgrid are thanked for research funding and computing facilities.

\end{document}